# New method for evaluating integrals involving orthogonal polynomials: Laguerre polynomial and Bessel function example

A. D. Alhaidari

*Shura Council, Riyadh 11212, Saudi Arabia*
AND
*Physics Department, King Fahd University of Petroleum & Minerals, Dhahran 31261, Saudi Arabia*
email: haidari@mailaps.org

**Abstract**: Using the theory of orthogonal polynomials, their associated recursion relations and differential formulas we develop a method for evaluating new integrals. The method is illustrated by obtaining the following integral result that involves the Bessel function and associated Laguerre polynomial:

$$\int_0^\infty x^\nu e^{-x/2} J_\nu(\mu x) L_n^{2\nu}(x)\,dx = 2^\nu \Gamma\!\left(\nu+\tfrac{1}{2}\right) \frac{1}{\sqrt{\pi \mu}} (\sin\theta)^{\nu+\frac{1}{2}} C_n^{\nu+\frac{1}{2}}(\cos\theta),$$

where $\mu$ and $\nu$ are real parameters such that $\mu \geq 0$ and $\nu > -\tfrac{1}{2}$, $\cos\theta = \frac{\mu^2 - 1/4}{\mu^2 + 1/4}$, and $C_n^\lambda(x)$ is the Gegenbauer (ultra-spherical) polynomial.

**Keywords**: Definite integral, Bessel function, Associated Laguerre polynomial, Gegenbauer polynomial, recursion relation.

The use of spherical Bessel functions[+] in the theoretical physics literature is overwhelming. One reason is that these functions are eigen-solutions of the three dimensional free wave operator (the Laplacian) in spherical coordinates [1]. Another is the fact that they make up the radial component of the wave functions for free particles in three dimensions [1,2]. These free-particle wave functions are the reference in scattering calculations for interactions with spherical symmetry. Consequently, the projection of the Bessel function onto various bases used in different numerical schemes is important. These weighted projections (typically, integrals) are necessary for computing the expectation values of selected observables to be compared with measurements. All integral formulas having the general form shown in the Abstract, which one could find in various tables of integration, are with the integration variable $x$ being replaced by $\sqrt{x}$ in the argument of the Bessel function [3]. This is useful for many applications where the Gaussian bases (a.k.a. "oscillator bases") are widely used in the calculation [4]. The elements of such a basis could be written as $x^\nu e^{-x^2/2} L_n^\lambda(x^2)$, or a linear combination thereof. Nonetheless, an equally useful basis, which is sometimes referred to as the "Laguerre basis", has elements of the form $x^\nu e^{-x/2} L_n^\lambda(x)$. In typical scattering problems with $x$ being the radial coordinate and $\nu = \ell + 1$, where $\ell$ is the angular momentum quantum number, the former basis is orthogonal whereas the latter is not; it is "trithogonal" (i.e., the basis overlap matrix – the identity representation – is tridiagonal). In this Letter, we use the theory of orthogonal polynomials, their associated recursion relations and differential formulas to develop a method for evaluating a new integral that involves the Bessel function $J_\nu(x)$ and the associated Laguerre polynomials $L_n^{2\nu}(x)$. In all subsequent developments we restrict our treatment to *real* spaces and fields.

---

[+] If $J_\nu(x)$ is the Bessel function, then the spherical Bessel function is defined as $j_\nu(x) = \sqrt{\frac{\pi}{2x}} J_{\nu+\frac{1}{2}}(x)$.

Since $J_\nu(x)$ is defined on the positive real line and the $\lim_{x\to 0} J_\nu(x) = \left(\frac{x}{2}\right)^\nu \big/ \Gamma(\nu+1)$ then we could, in principle, expand it as an infinite series in terms of a complete set of basis functions which are compatible with the range of $J_\nu(x)$ and its limiting values. Therefore, we assume that we can make a separable expansion of $J_\nu(\mu x)$ in the space spanned by the square integrable functions $\{\phi_n(x)\}_{n=0}^\infty$ with real expansion coefficients $\{c_n(\mu)\}_{n=0}^\infty$. That is, we set out to find real functions $\{c_n(\mu)\}_{n=0}^\infty$ such that

$$J_\nu(\mu x) = \sum_{n=0}^\infty c_n(\mu) \phi_n(x) = \sum_{n=0}^\infty c_n(\mu) x^\nu e^{-\alpha x} L_n^\lambda(x), \qquad (1)$$

where $\mu \geq 0$, $\alpha > 0$, and $\lambda > -1$. Using the orthogonality property of the Laguerre polynomials [5] we can write

$$c_n(\mu) = \frac{\Gamma(n+1)}{\Gamma(n+\lambda+1)} \int_0^\infty x^{\lambda-\nu} e^{(\alpha-1)x} J_\nu(\mu x) L_n^\lambda(x) \, dx. \qquad (2)$$

Therefore, we must impose the stronger constraint that $1 > \alpha > 0$. The differential equation for the Bessel function could be written as the eigenvalue equation $\mathcal{D} J_\nu(x) = -J_\nu(x)$, where $\mathcal{D}$ is the second order linear differential operator $\frac{d^2}{dx^2} + \frac{1}{x}\frac{d}{dx} - \frac{\nu^2}{x^2}$ [5]. Using the differential equation of the Laguerre polynomials and their differential formula, $x\frac{d}{dx} L_n^\lambda = n L_n^\lambda - (n+\lambda) L_{n-1}^\lambda$, we can write

$$\mathcal{D}\phi_n = \left(n\frac{2\nu-\lambda}{x^2} - \alpha\frac{2n+2\nu+1}{x} + \alpha^2\right)\phi_n - \left(\frac{2\nu-\lambda}{x^2} + \frac{1-2\alpha}{x}\right)(n+\lambda)\phi_{n-1}. \qquad (3)$$

We define a conjugate space spanned by the real $L^2$ functions $\{\tilde{\phi}_n(x)\}_{n=0}^\infty$ such that the basis overlap matrix (i.e., the inner product $\langle\tilde{\phi}_n|\phi_m\rangle$ with the integration measure $dx$) is tridiagonal. That is, $\langle\tilde{\phi}_n|\phi_m\rangle = 0$ if $|n-m| \geq 2$. Thus, if we write $\tilde{\phi}_n(x) = x^\rho e^{-\beta x} L_n^\lambda(x)$, then the orthogonality of the Laguerre polynomials dictates that

$$\beta = 1-\alpha \text{ and } \rho = -\nu + \lambda + \delta, \qquad (4)$$

where $\delta = 0$ or $1$. Now, the matrix representation of the eigenvalue equation $\mathcal{D} J_\nu(\mu x) = -\mu^2 J_\nu(\mu x)$ could be written as

$$\langle\tilde{\phi}_n|(\mathcal{D} + \mu^2)|J_\nu\rangle = \sum_{m=0}^\infty c_m \langle\tilde{\phi}_n|(\mathcal{D} + \mu^2)|\phi_m\rangle = 0. \qquad (5)$$

Additionally, we also require that the matrix representation of the differential operator $\mathcal{D}$ be at most tridiagonal. That is, $\langle\tilde{\phi}_n|\mathcal{D}|\phi_m\rangle = 0$ if $|n-m| \geq 2$. Using the action of $\mathcal{D}$ on the basis given by Eq. (3) and the parameter relation (4) this requirement translates into $\lambda = 2\nu$ and $\delta = 1$. Hence, we could write

$$\mathcal{J}_{nm} \equiv \langle\tilde{\phi}_n|(\mathcal{D} + \mu^2)|\phi_m\rangle = (\alpha^2 + \mu^2)\langle\tilde{\phi}_n|\phi_m\rangle - \alpha(2m+2\nu+1)\langle\tilde{\phi}_n|\tfrac{1}{x}|\phi_m\rangle \\ + (2\alpha-1)(m+2\nu)\langle\tilde{\phi}_n|\tfrac{1}{x}|\phi_{m-1}\rangle \qquad (6)$$

Again, using the orthogonality relation of the Laguerre polynomials we obtain

$$\mathcal{J}_{nm} = \frac{\Gamma(n+2\nu+1)}{\Gamma(n+1)}\Big[(2n+2\nu+1)(\alpha^2+\mu^2-\alpha)\delta_{nm} - n(\alpha^2+\mu^2)\delta_{n,m+1} \\ -(n+2\nu+1)(\alpha^2+\mu^2+1-2\alpha)\delta_{n,m-1}\Big] \qquad (7)$$

Therefore, Eq. (5) results in a three-term recursion relation for the expansion coefficients which is written as $\mathcal{J}_{n,n-1}c_{n-1} + \mathcal{J}_{n,n}c_n + \mathcal{J}_{n,n+1}c_{n+1} = 0$. In terms of the polynomials $\{P_n\}_{n=0}^\infty$, which are defined as

$$P_n(\mu) \equiv \frac{\Gamma(n+2\nu+1)}{\Gamma(n+1)} c_n(\mu), \tag{8}$$

this recursion relation reads as follows

$$(2n+2\nu+1)(\alpha^2+\mu^2-\alpha)P_n = (n+1)(\alpha^2+\mu^2+1-2\alpha)P_{n+1} \\ +(n+2\nu+1)(\alpha^2+\mu^2)P_{n-1} \tag{9}$$

The two-parameter polynomial, $P_n^{\alpha,\nu}(\mu)$, which satisfies this recursion relation is, to the best of our knowledge, new. However, for $\alpha = \tfrac{1}{2}$ Eq. (9) reduces to the three-term recursion relation of the Gegenbauer (ultra-spherical) polynomial $C_n^{\nu+\tfrac{1}{2}}(y)$ [5], where $y = y(\mu) = \frac{\mu^2 - 1/4}{\mu^2 + 1/4} \equiv \cos\theta$ and $0 < \theta \leq \pi$. Consequently, we can write

$$P_n(\mu) = f_\nu(\mu) C_n^{\nu+\tfrac{1}{2}}(y), \tag{10}$$

where $f_\nu(\mu)$ is an arbitrary real function of $\mu$ which is independent of the index $n$. Combining Eq. (2) with Eq. (8) we get

$$P_n(\mu) = \int_0^\infty x^\nu e^{-x/2} J_\nu(\mu x) L_n^{2\nu}(x)\, dx = f_\nu(\mu) C_n^{\nu+\tfrac{1}{2}}(y). \tag{11}$$

Therefore, what remains is only to evaluate $f_\nu(\mu)$. To do that, we differentiate the above integral with respect to $\mu$ using the chain rule (in the sequence $\mu \to \mu x \to x$) and then integrate by parts since the integrand vanishes at the end points. The result is as follows:

$$\mu \frac{dP_n}{d\mu} = -\int_0^\infty x^\nu e^{-x/2} J_\nu(\mu x) \left[ (\nu+1) L_n^{2\nu} - \tfrac{x}{2} L_n^{2\nu} + x \tfrac{d}{dx} L_n^{2\nu} \right] dx \\ = -\tfrac{1}{2}\int_0^\infty x^\nu e^{-x/2} J_\nu(\mu x) \left[ L_n^{2\nu} + (n+1)L_{n+1}^{2\nu} - (n+2\nu)L_{n-1}^{2\nu} \right] dx \tag{12}$$

In going from the first to the second line of this equation, we have used the differential formula and recursion relation for $L_n^{2\nu}(x)$. Thus, Eq. (12) states that

$$2\mu \frac{dP_n}{d\mu} = -P_n - (n+1)P_{n+1} + (n+2\nu)P_{n-1}. \tag{13}$$

On the other hand, differentiating the right side of Eq. (11) with respect to $\mu$ and using $\mu \tfrac{d}{d\mu} = (1-y^2)\tfrac{d}{dy}$ in the differential relation of the Gegenbauer polynomial,

$$(1-y^2)\frac{d}{dy} C_n^\lambda(y) = -\frac{1}{2}\frac{n(n+1)}{n+\lambda} C_{n+1}^\lambda(y) + \frac{1}{2}\frac{(n+2\lambda-1)(n+2\lambda)}{n+\lambda} C_{n-1}^\lambda(y), \tag{14}$$

we obtain

$$2\mu \frac{dP_n}{d\mu} = \left(2\mu \frac{df_\nu}{d\mu}\right) C_n^{\nu+\tfrac{1}{2}} - \frac{n(n+1)}{n+\nu+\tfrac{1}{2}} P_{n+1} + \frac{(n+2\nu)(n+2\nu+1)}{n+\nu+\tfrac{1}{2}} P_{n-1}. \tag{15}$$

Equating the right hand side of Eq. (13) with that of Eq. (15) and using the recursion relation of the Gegenbauer polynomial we conclude that

$$\left(\mu \frac{d}{d\mu} + \tfrac{1}{2}\right) f_\nu(\mu) = -\left(\nu + \tfrac{1}{2}\right) y\, f_\nu(\mu). \tag{16}$$

Defining $g_\nu(y) = \sqrt{\mu} f_\nu(\mu)$, this could be rewritten for $g_\nu$ as $(1-y^2)\frac{d}{dy} g_\nu = -\left(\nu + \frac{1}{2}\right) y\, g_\nu$ giving the solution $f_\nu(\mu) = A_\nu \frac{1}{\sqrt{\mu}} (\sin\theta)^{\nu+\frac{1}{2}}$, where $A_\nu$ is a constant which is independent of $\mu$ and $\sin\theta = \frac{\mu}{\mu^2+1/4}$.

Putting all of the above together, we obtain the following realization of the real expansion coefficients $c_n(\mu)$:

$$c_n(\mu) = \frac{\Gamma(n+1)}{\Gamma(n+2\nu+1)} \frac{A_\nu}{\sqrt{\mu}} (\sin\theta)^{\nu+\frac{1}{2}} C_n^{\nu+\frac{1}{2}}(\cos\theta). \qquad (17)$$

To determine $A_\nu$ we substitute this in the expansion (1) and take the simultaneous limit $x \to 0$ and $\mu \to 0$. Using $C_n^\lambda(-1) = (-1)^n \frac{\Gamma(n+2\lambda)}{\Gamma(n+1)\Gamma(2\lambda)}$, we obtain

$$\left(\frac{\mu x}{2}\right)^\nu \frac{1}{\Gamma(\nu+1)} = A_\nu (\mu x)^\nu 2^{2\nu+1} \frac{1}{\Gamma(2\nu+1)} \sum_{n=0}^\infty (-1)^n L_n^{2\nu}(0). \qquad (18)$$

Using the generating function $\sum_{n=0}^\infty L_n^\lambda(x) t^n = (1-t)^{-1-\lambda} \exp\left(\frac{xt}{t-1}\right)$, which is valid for $-1 \le t < 1$ [6], we get the following value for the constant $A_\nu = \frac{1}{\sqrt{\pi}} 2^\nu \Gamma\left(\nu + \frac{1}{2}\right)$. Inserting this in the expression $f_\nu(\mu) = A_\nu \frac{1}{\sqrt{\mu}} (\sin\theta)^{\nu+\frac{1}{2}}$ and substituting in Eq. (11) we, finally, get the sought-after result

$$\int_0^\infty x^\nu e^{-x/2} J_\nu(\mu x) L_n^{2\nu}(x)\, dx = 2^\nu \Gamma\left(\nu + \frac{1}{2}\right) \frac{1}{\sqrt{\pi\mu}} (\sin\theta)^{\nu+\frac{1}{2}} C_n^{\nu+\frac{1}{2}}(\cos\theta). \qquad (19)$$

In conclusion, we summarize the method in the following steps:
1) We write $J_\nu(\mu x)$ as an infinite series of products of functions, $\{\phi_n(x)\}_{n=0}^\infty$ and $\{c_n(\mu)\}_{n=0}^\infty$, that are compatible with the range of definition of $J_\nu(x)$ and its limit values: Eq. (1)
2) Using the orthogonality property of the polynomials in the basis functions $\{\phi_n\}$ we write an integral expression for the expansion coefficients $\{c_n\}$: Eq. (2)
3) We construct the tridiagonal conjugate space spanned by $\{\tilde{\phi}_n(x)\}_{n=0}^\infty$ and require that it also supports a tridiagonal matrix representation for the Bessel differential operator $\mathcal{D}$: Eq. (6)
4) The resulting three-term recursion relation is solved in terms of orthogonal polynomials giving the expansion coefficients $\{c_n\}$ modulo an arbitrary real function $f_\nu(\mu)$: Eq. (10)
5) The function $f_\nu(\mu)$ is determined (up to an overall constant factor $A_\nu$) by solving a simple first order linear differential equation which is obtained by differentiating the resulting integral formula: Eqs. (11-15)
6) The remaining constant $A_\nu$ is determined by taking the limit of the expansion of $J_\nu(\mu x)$ as $(\mu, x) \to 0$: Eq. (18)

One should also note that we needed knowledge of only two properties of the Bessel function to obtain the analytic closed-form result. These were the differential equation of $J_\nu(x)$ and its values at the limits. This is a highly nontrivial observation that could have a major impact on the application of the method on a wider range of functions. It adds value to the method and could motivate its development into a powerful integration tool for such problems.

**Acknowledgments**: The author is grateful to H. A. Mavromatis (Physics, KFUPM) and R. S. Alassar (Math, KFUPM) for fruitful consultations. Help in the calculation provided by A. Al-Hasan (Physics, KFU) is highly appreciated.

**References:**

[1] P. M. Morse and H. Feshbach, *Methods of Theoretical Physics*, Vol. I (McGraw-Hill, New York, 1953); R. Courant and D. Hilbert, *Methods of Mathematical Physics*, Vol. I (Interscience, New York, 1966); G. Arfken, *Mathematical Methods for Physicists*, 2nd ed. (Academic, New York, 1970) pp. 478-533

[2] A. Messiah, *Quantum Mechanics*, Vol. I (North-Holland, Amsterdam, 1965); E. Merzbacher, *Quantum Mechanics*, 2nd ed. (Wiley, New York, 1970); R. L. Liboff, *Quantum Mechanics*, 2nd ed. (Addison-Wesley, Reading, 1992)

[3] See, for example, I. S. Gradshteyn and I. M. Ryzhik, *Tables of Integrals, Series, and Products* (Academic, New York, 1980)

[4] R. G. Newton, *Scattering Theory of Waves and Particles*, 2nd ed. (Springer, New York, 1966); S. Gelman, *Topics in Atomic Collision Theory* (Academic, New York, 1969); B. H. Bransden, *Atomic Collision Theory* (Benjamin, New York, 1970); J. R. Taylor, *Scattering Theory* (Wiley, New York, 1972); P. G. Burke, *Potential Scattering in Atomic Physics* (Plenum, New York, 1977)

[5] A. Erdélyi (Ed.), *Higher Transcendental Functions*, Vol. I (McGraw-Hil1, New York, 1953); M. Abramowitz and I. A. Stegun (Eds.), *Handbook of Mathematical Functions* (Dover, New York, 1965); T. S. Chihara, *An Introduction to Orthogonal Polynomials* (Gordon and Breach, New York, 1978); G. Szegö, *Orthogonal polynomials*, 4th ed. (Am. Math. Soc., Providence, RI, 1997)

[6] W. Magnus, F. Oberhettinger, and R. P. Soni, *Formulas and Theorems for the Special Functions of Mathematical Physics*, 3rd ed. (Springer, New York, 1966) pp. 242